\documentclass[%
 reprint,
 amsmath,amssymb,
 aps,
]{revtex4-1}

\usepackage[dvipdfmx]{graphicx}
\usepackage{dcolumn}
\usepackage{bm}

\newcommand{\cf}{$f_0$\hspace{1mm}}
\newcommand{\Qinv}{$Q^{-1}$\hspace{1mm}}
\newcommand{\Q}{$Q$\hspace{1mm}}

\begin{document}


\title{Current and  Magnetic Field Dependences of a Superconducting Coplanar Waveguide Resonator}

\author{H. Kurokawa}
 \email{kurokawa@maeda1.c.u-tokyo.ac.jp}
 \author{F. Nabeshima}
 \author{A. Maeda}
\affiliation{%
 Department of Basic Science, The University of Tokyo, 3-8-1, Komaba, Meguro-ku, Tokyo 153-8902, Japan
}%

\date{\today}

\begin{abstract}
We fabricated superconducting coplanar waveguide resonator with leads for dc bias, which enables the ac conductivity measurement under dc bias. The current and the magnetic field dependences of resonance properties were measured, and hysteretic behavior was observed as a function of the dc driving current. The observed shift in the inverse of the quality factor, \Qinv, and the center frequency, \cf, were understood by considering both the motion of vortices and the suppression of the order parameter with dc current. Our investigation revealed that the strongly pinned vortices have little infuluence on the change in \cf, while it still affects that of \Qinv. Our results indicate that an accurate understanding of the dynamics of driven vortices is indispensable when we attempt to control the resonance properties with high precision.
\begin{description}
\item[PACS numbers]74.25.Ha, 84.40.Dc, 03.67.Lx

\end{description}
\end{abstract}

\maketitle


\section{\label{sec:intro}Introduction}
Superconducting coplanar waveguide resonators are becoming essential components in advanced measurement devices. They are integrated into various circuits for the manupulation of quantum bits
\cite{Sillanpaa2007
	,Clarke2008
}, used for studies of the circuits quantum electrodynamics
\cite{Blais2004
	,Wallraff2004
	,You2011
 }
 and also for the detction of the photon from the space\cite{Mazin2002,Day2003,Zmuidzinas2004}. For further advanced applications and improving its versatility, various techniques for manipulating the resonance characteristics have been developed
 \cite{Osborn2007
	,Sandberg2008
	,Healey2008
	,Chen2011
	,Li2013
	,Hao2014
}. 
\par
Indroducing dc bias current to the resonator or applying magnetic field were suggested to tune the resonance properties, and how they work have been widely investigated. Indeed, dc bias current decreases the center frequency, \cf, of the resonator because of the suppression of the order parameter of the superconductor\cite{Gittleman1965,Lofgren1997}. In order to control \cf by dc bias current keeping the high quality factor, \Q,  several architectures were proposed\cite{Chen2011,Li2013}. For instance, Chen \textit{et al}.　proposed a half wave length filter\cite{Chen2011}, and they investigated the effect of dc bias on the resonance property. Li and Kycia fabricated more adaptable $\lambda/4$ filter incorporated into the resonator\cite{Li2013}. Meanwhile, operating the resonator under the magnetic field is required in hybrid systems consisting of ultracold atoms, molecules, or a single electron and the superconducting resonator\cite{Verdu2009
	,Rabl2006
   ,Bushev2011
}
. Applying magnetic field causes additional loss due to the vortices penetrating into the resonator, and it induces hysteresys in the magnetic field dependence of \cf and \Q\cite{Healey2008,Song2009,Bothner2012}, resulting in the performance degradation of the resonator. For minimizing those undesireble effects, fabrication of a slot or antidots was proposed, and  it was confirmed to reduce the loss by trapping the vortices inside it\cite{
	Song2009
}.　In addition, vortices are subject to the driving force from the current, so that resonance properties would show complex behavior if we operate the resonator under both the driving current and the magnetic field.
 However, to the best of our knowledge, there are no reports about the resonance characteristics with the presence of both dc bias current and magnetic field. Since Vortices show complex dynamics due to the interaction with the pinning and also by the driving current
 \cite{Gittleman1968
    ,Campbell1972a
}, understandings of the influence of dc bias current on the resonator under magnetic field is a physically interesting problem as well as its importance for applications. Thus, we investigated the current and the magnetic-field dependence of the \Q and \cf of a superconducting resonator.
\par
 In this letter, we fabricated the superconducting coplanar waveguide resonator with leads for the introducion dc bias current.  The dc current dependence of \cf and \Qinv was measured with and without magnetic field perpendicular to the resonator. Hysteretic behaviour was observed in the 1st and the 2nd current sweep with the magnetic field. In particular, the behavior of \Qinv was complicated, in contrast to that of \cf. These behaviour were explained considering the motion of vortices under the pinning.  Our results reconfirmed the importance of considering the dynamics of vortices for the  precise control of the resonance characteristics of the superconducting resonator.
 \section{\label{sec:meth}Experimental method}
 \begin{figure}[th]
 	\centering
 	\includegraphics[width=0.9\linewidth]{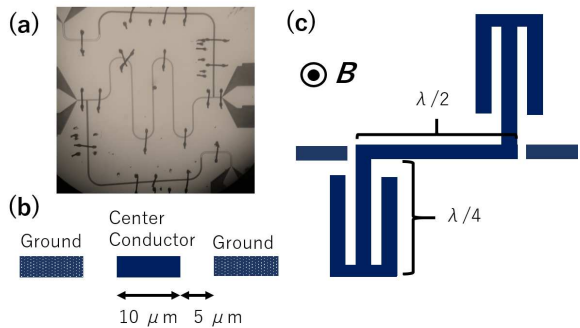}
 	\caption{Coplanar waveguide resonator with dc leads.(a)The optical image. (b)The cross section. (c)A schematic representation of the resonator pattern.}
 	\label{fig:resonator}
 \end{figure}
 Fig.\ref{fig:resonator} shows our superconducting coplanar transmission line 
 $\lambda/2$ resonator whose center frequency was designed to be 8 GHz. The total length of the resonator was 8 mm. Fig.\ref{fig:resonator}(b) shows the cross section of the resonator and the filter. It was desined so that the characteristic impedance becomes 50 $\Omega$. The resonator was capacitively coupled to the input and output feeds through a $5\  \mu\textrm{m}$ gap, and it had extra leads at the ends for introdung the dc bias current. The length of the lead was  equal to be $\lambda/4$, and there were also $\lambda/4$ open stubs at the root of the lead. Thus, it serves as a $\lambda/2$ bandstop filter, which reduces the couplings to the external circuits, maintaining relatively high \Q\cite{Li2013}. In fact, we realized $Q(0 \ \textrm{mA}, 0\ \textrm{G})$ of 9000 despite the attachment of the dc leads. 
 \par  The resonator was made of Nb, which was deposited on 5 mm $\times$ 5 mm $\times$ 0.5 mm MgO substrate by rf sputtering. Deposition rate was monitored by the quatz oscillator and the thickness was estimated to be 100 nm. Transition temperature of the film was 8.7 K and resiudal resistivity was 7 $\mu\Omega $cm. In addition, we measured the dc $I-V$ characteristics with the standard dc four probe method, and the critical current was 70 mA at 3 K, at 30 G, as shown in Fig.\ref{fig:171118nb318gs213k30graw}(a). Then, the structure of the resonator was patterned onto the film by the standard photolithography technique. The patterned film was etched by　the reactive ion etching using $\textrm{CF}_4$ gas. 
 \par  The fabricated resonator was connected to a printed circuit board with alminum bonding wires, and the circuit board was attached to the semirigid coaxial cables with subminiture-A connectors and subminiture-P connectors. The device was inserted into PPMS (Quantum Design), and was cooled down to 3 K. Magnetic field was applied perpendicular to the film. All the measurements were done under the field-cooled conditions.

 We measured the resonance spectrum with a network analyzer N5222A(Keysight), with increasing dc bias current, and thus obtained the current dependence of \cf and \Qinv. In order to minimize the effect of joule heating, the dc current was introduced as a rectangular pulse whose width was 25 ms and duty cycle was 0.1. A function generator was used to generate the recutangular  pulse and trigger the network analyzer. The amplitude of the current was estimated from the voltage drop at a 10 $\Omega$ resistance in series with the resonator. It was measured by an oscilloscope beforehand the measurements.
  \section{\label{sec:resu}Results}
\begin{figure}[b]
	\centering
	\includegraphics[width=1\linewidth]{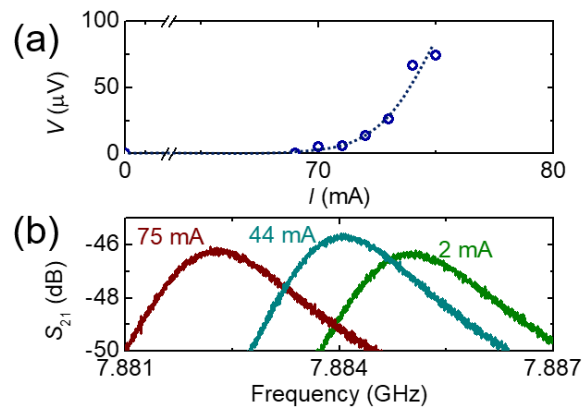}
	\caption{(a) The $I-V$ characteristics of the Nb film measured at 3 K, 30 G. The blue dots are the guide for eye. (b) The resonance spectrum of the 7.9 GHz resonator measured with increasing dc current.}
	\label{fig:171118nb318gs213k30graw}
\end{figure}
Fig.\ref{fig:171118nb318gs213k30graw}(b) shows the resonance spectra measured with increasing dc bias current at 3 K, and at 30 G. The current dependence of the spectrum was clearly seen. The asymmetry in the spectrum may be originated from parasitic capacitances at the coupling feeds and the dc leads\cite{Hornibrook2012}. We fitted these spectra by a lorentzian and obtained the change in \Qinv  and \cf as a function of the dc current. \Qinv and \cf were also found to show hysteresis, depending on the current history.
\par
\begin{figure}[t]
	\centering
	\includegraphics[width=0.8\linewidth]{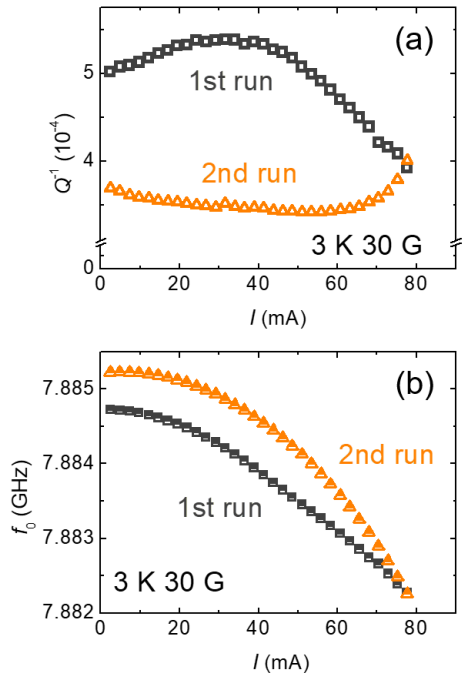}
	\caption{The current dependence of (a) \Qinv and (b) \cf at 3 K, at 30 G. Gray square is the data in the 1st run, orange triangles are those in the 2nd run.}
	\label{fig:1st2nd}
\end{figure}

Fig.\ref{fig:1st2nd}(a)  and Fig.\ref{fig:1st2nd}(b) show the current dependences of \Qinv and \cf, respectively. We repeated the measurement cycle more than twice, but the current dependence was unchanged after the 2nd run. Therefore, we plotted only the results in the 1st and the 2nd run.  In the 1st run, \Qinv gradully increased with the dc current until 35 mA and then decreased, while \cf  monotonically decreased. On the other hand, in the 2nd run, \Qinv  slightly decreased at the lower current regime and increased over 60 mA, while \cf  continued to decrease just like in the 1st run. In addition, by comparing the value of 2 mA between the 1st run and the 2nd run, \Qinv  was smaller by $1.3\times10^{-4}$ and \cf  was larger by 0.5 MHz at the 2nd run.
\par
\begin{figure}[tbh]
	\centering
	\includegraphics[width=0.8\linewidth]{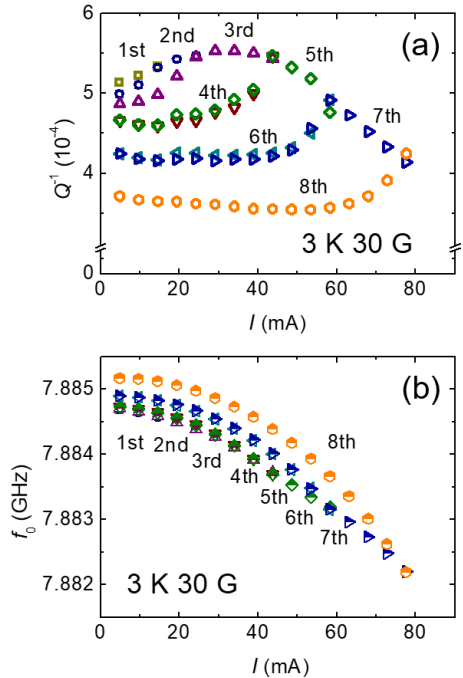}
	\caption{The current dependence of (a) \Qinv and (b) \cf measured with increasing maximum dc current at 3 K, at 30 G.}
	\label{fig:sbs}
\end{figure}

In order to understand the current dependence and the hysteretic behaviour, we measured the change of \Qinv  and \cf by gradually increasing maximam dc current, as shown in Fig.\ref{fig:sbs}.
Fig.\ref{fig:sbs}(a) shows the change of \Qinv. Until in the 3rd run, \Qinv increased almost in the same manner except for small differences between in the 2nd run and in the 3rd run. In the 4th run, however, the current dependence showed a remarkable difference. The increase was moderate under 30 mA when compared with the data in the 3rd run, then increased to the value of the 3rd run at 44 mA. In the 5th run, the data showed the same current dependence as in the 4th run under 44 mA, then started to decrease. The changes in the 6th run and in the 7th run resemble those in the 4th run and in the 5th run, respectively. The 8th run corresponded to the 2nd run of Fig.\ref{fig:1st2nd}(a). On the other hand, the changes in \cf  shown in Fig.\ref{fig:sbs}(b) was rather subtle. \cf  always decreased with increasing dc current and no dramatic change was observed around 40 mA and 60 mA, where \Qinv  started to decrease or increase. As we increase the maximum dc current, the initial value of \cf  gradually increased, which was similar to those shown in Fig.\ref{fig:1st2nd}(b). Additionally, the 8th run also corresponded to the 2nd run of Fig.\ref{fig:1st2nd}(b) just like \Qinv.
 \section{\label{sec:dis}Discussions}
We first consider the complicated behavior observed in the current dependence and the hysteresis in \Qinv, and then discuss both of those in \cf, in terms of vortices. The change in \Qinv  of Fig.\ref{fig:1st2nd}(a) can be roughly classified into three types; (1) the increase under 40 mA, (2) the decrease over 40 mA in the 1st run, and (3) the increase over 60 mA in the 2nd run. According to Fig.\ref{fig:171118nb318gs213k30graw}(a), no macroscopic motion of vortices occurs under 70 mA, because there was no dc electric field induced by the translational motion of vortices. Hence, the shift under 70 mA is irrelevant to the translational motion of vortices, and it should originate from the local changes of the configuration of the vortex lattice due to the dc driving current. On the other hand, the slight increase of \Qinv observed around 70 mA in the 2nd run may be related to the macroscopic translational motion of vortices. 
\par
Considering the local motion of vortices, the increase and the decrease in the 1st run can be explained as follows. The change in \Qinv is directly related to the change of the resistance of the resonator as $\Delta R = \Delta(Q^{-1})/\omega L$. Several factors contributes to \Qinv: the diectric loss, $Q_\textrm{die}$, the conductor loss, $Q_\textrm{c}$, the radiation loss, $Q_\textrm{rad}$, the loss in the external circuits, $Q_\textrm{ex}$, and the loss due to the vortices, $Q_\textrm{v}$. Assuming that the dc bias current affects only on  the vortices, the changes in \Qinv is equal to the changes in the resistivity of the vortices.  The increase of \Qinv means that the ac response of vortices became more resistive. This indicates that the vortices felt weaker pinning force because of the dc current. If we continued to increase the dc current, the vortices could gradually move to more stable, strongly pinned sites. Thus, the strongly pinned vortices responsed to the ac current less resistively, resulting in the decrease of \Qinv above 40 mA. 
\par
In contrast, as was mentioned earlier, the increase of \Qinv in the 2nd run may results from the translational motion of vortices. Such motions reduced the pinning force, leading to the more resistive response to the ac current. It is worth noting that simple models of the ac response can not explain this gradual increase. Gittleman and Rosenblum formulated the ac response of the vortex on the basis of the phenomenological equation of motion of the vortex: $\eta v + kx = j_{\textrm{ac}}\phi_0$, where $\eta$ is the viscous drag coefficient, $\phi_0$ is the flux quantum\cite{Gittleman1968}. The pinning constant, $k$, is the curvature of the pinning potential, $U_{\textrm{p}}$. When we apply the dc critical current, this model predicts that $U_{\textrm{p}}$ and $k = \textrm{d}^2U/\textrm{d}x^2$ becomes zero, since $U_{\textrm{p}}\rightarrow0$ at the critical current. Vanishing of the pinning potential should lead to purely resistive response, i.e. the flux flow. Shklovskij and Dobrovolskkiy also showed a sharp increase of resistivity at the dc critical current by numerical calculation of the equation of motion\cite{Shklovskij2008}. Instead, what we observed is the gradual increase rather than such a drastic sudden increase around the dc critical current. More elaborate theory is indispensable for further understandings of the pinning and the ac response of the dc driven vortices.
\par
In addition to the changes mentioned above, we briefly discuss the subtle decrease of \Qinv in the 2nd run. We considered that  the unharmonicity in $U_{\textrm{p}}$ can explain this decrease. If $U_{\textrm{p}}$ was harmonic everywhere, $k$ would be unchanged irrespective with the position of the votex, and \Qinv would be independent of the dc current. On the other hand, if the potential is unharmonic\cite{Prozorov2003}, $k$ depends on the displacement of the vortex. In such cases, $k$ gets larger and the pinning force becomes stronger as the vortex moves from its initial position. Stronger pinning force reduces the ac resistance of the vortex, leading to the decrease of \Qinv in the 2nd run.
\begin{figure}[tb]
	\centering
	\includegraphics[width=0.8\linewidth]{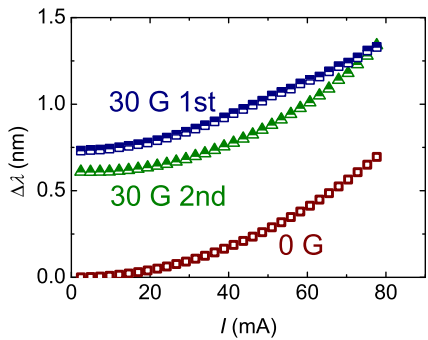}
	\caption{The current dependence of the change in the penetration depth at 0 G, in the 1st run at 30 G, and in the 2nd run at 30 G.}
	\label{fig:171118dellambda0g30g}
\end{figure}
\par
Next, we consider the change in \cf. As we increase the dc current, \cf continued to decrease parabolically, and it became larger by 0.5 MHz after the current sweep. We consider that the paraboric current dependence of the change in \cf is due to the current dependence of the order parameter of the superconductor\cite{Gittleman1965}.  From Ginzburg-Landau equation, in thin films, current supress the order parameter as $\lvert\psi\rvert^2=1-Aj^2$, where $\psi$ is the order parameter, $A$ is a constant which depends only on temperature. Imaginary part of the complex conductivity, $\sigma_2$, is proportional to $\lvert\psi\rvert^2$. The change in inductance, $\Delta L$, can be written as $\Delta L\propto\sigma_2(j)-\sigma_2(0)\propto j^2$, and $\Delta L=-2(\Delta f/f_0)L$. Hence, the shift in the center frequency, $\Delta f$, turns out to be proportional to the square of the current density, which results in the parabolic shift in \cf.
\par
To explain the hysteretic behavior in \cf, we calculated the change of the penetration depth, $\Delta\lambda$. According to Watanabe\cite{Watanabe1994}, 
total inductance, $L$, of the resonator is the sum of the magnetic inductance, $L_\textrm{m}$, and the kinetic inductance of a superconductor, $L_{\textrm{k}}$, which can be expressed as $L_{\textrm{k}}=\mu_0(\lambda^2/dw)g$, where $d$ and $w$ are the thickness and the width of the center conductor, respectively, and $g$ is the geometric factor. Then, we obtained $\Delta\lambda=(dw/2\lambda g \mu_0)\Delta L_{\textrm{k}}$.  We calculated $L(0\ \textrm{mA},\ 0\ \textrm{G})$ with the assumption that $\lambda(0\ \textrm{mA},\ 0\ \textrm{G})$ is 96 nm using the relations, $\lambda = 1.05\sqrt{\rho(T_\textrm{c})/T_\textrm{c}}\times10^{-3}\ $m\cite{Orlando1979} and $\lambda(t)=\lambda_0/\sqrt{1-t^4}$. Fig.6 shows the current dependence of $\Delta\lambda$ at 0 G, and that in the 1st run at 30 G, and that in the 2nd run at 30 G. When temperature, $T$, is not so close to the transition temperature, $T_\textrm{c}$, and also when the magnetic field, $B$, is much lower than the upper critical field, $B_\textrm{c2}$,  $\lambda^2$ becomes a sum of the superconducting penetration depth, $\lambda_s$, and the vortex(campbell) penetration depth, $\lambda_v$:  $\lambda^2=\lambda_s^2+\lambda_v^2$ \cite{Coffey1991,Golosovsky1996}.  We performed measurements $T/T_\textrm{c}\sim0.35$ and $B/B_\textrm{c2}\sim2\times10^{-4}$, so that the condition is fulfilled. Considering that  $\lambda_s$ remains almost unchanged at this low magnetic field, the difference in $\Delta\lambda$(2 mA) between 0 G and 30 G is mainly due to the vortices. $\Delta\lambda$ of 0 G and that of the 2nd run at 30 G showed the same dependence on dc current, indicating that only the suppression of the $\psi_0$ dominates the $\Delta\lambda$, and vortices do not contribute to the change. On the other hand, the 1st run at 30 G showed a slight different behavior comparing with the 2nd run. This difference should come from the difference in $\lambda_v$, resulting from the local motion of the vortices. In the 2nd run, $\lambda_v$ is shorter than that in the 1st run, suggesting that the vortices are more strongly pinned in the 2nd run. It is consistent with the result of \Qinv. 
 \section{\label{sec:conc}Conclusions}
We fabricated the superconducting coplanar waveguide resonator with leads for introducing dc bias, and measured the current and the magnetic field dependence of the resonance characteristics, \Qinv and \cf. We observed the current dependence of \Qinv and \cf together with the hysteresis. These are considered to originate from both the changes in the vortex configuration and the suppression of the order parameter. Applying dc current caused the local motion of the vortices, and they felt the weaker pinning force at first, then were gradually trapped to the more stable pinning sites. Therefore, \Qinv increased at the lower current regime, while it decreased at the higher current regime in the 1st run. Meanwhile, the current dependence of \cf was mainly due to the change of the order parameter of the superconductor. These resluts suggest that the influence of the vortex must be considered carefully for the accurate control of the resonance properties of the superconducting resonator under both dc bias and magnetic field.
\section*{Acknowledgements}
This work was supported by JSPS KAKENHI Grant Number JP16H00795.
\bibliographystyle{apsrev4-1.bst}
%


\end{document}